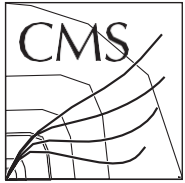
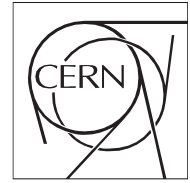

**The Compact Muon Solenoid Experiment**

# Conference Report

# The CMS Outer Tracker for the High-Luminosity LHC

Erik Butz for the CMS Tracker Group

**Abstract**

The era of High-Luminosity Large Hadron Collider will pose unprecedented challenges for detector design and operation. The planned luminosity of the upgraded machine is 5–7.5 × $10^{34}$ cm$^{-2}$s$^{-1}$, reaching an integrated luminosity of 3000–4000 fb$^{-1}$ by the end of 2039. The CMS Tracker detector will have to be replaced in order to fully exploit the delivered luminosity and cope with the demanding operating conditions. The new detector will provide robust tracking as well as input for the first level trigger. This report is focusing on the replacement of the CMS Outer Tracker system, describing new layout and technological choices together with some highlights of research and development activities.

# The CMS Outer Tracker for the High-Luminosity LHC

Erik Butz[a,*]

*for the CMS Tracker Group*

[a]*Karlsruhe Institute of Technology*

**Abstract**

The era of High-Luminosity Large Hadron Collider will pose unprecedented challenges for detector design and operation. The planned luminosity of the upgraded machine is 5–7.5 × $10^{34}$ cm$^{-2}$s$^{-1}$, reaching an integrated luminosity of 3000–4000 fb$^{-1}$ by the end of 2039. The CMS Tracker detector will have to be replaced in order to fully exploit the delivered luminosity and cope with the demanding operating conditions. The new detector will provide robust tracking as well as input for the first level trigger. This report is focusing on the replacement of the CMS Outer Tracker system, describing new layout and technological choices together with some highlights of research and development activities.

## 1. Introduction

With the Large Hadron Collider (LHC) in its 10$^{th}$ year of operation, CERN is already planning the next upgrade of its accelerator complex. The High-Luminosity LHC (HL-LHC) [1] is a luminosity upgrade of the LHC accelerator foreseen to be carried out during the years 2024 and 2025 during the *long shutdown 3* (LS3). After this upgrade the LHC will be providing instantaneous luminosities of initially 5 × $10^{34}$ cm$^{-2}$s$^{-1}$ ramping up to 7.5 × $10^{34}$ cm$^{-2}$s$^{-1}$. The number of simultaneous inelastic collisions per *pp* bunch crossing (referred to as *pile-up*) will be 140 (initial luminosity) to 200 (ultimate luminosity). The total integrated luminosity for the HL-LHC running period is expected to be up to 4 ab$^{-1}$. The currently installed detectors, notably the strip or outer tracker of CMS will not withstand the expected radiation dose and their performance will be inadequate to deal with the challenging environment posed by the collisions at the HL-LHC. Thus CMS foresees [2] a complete replacement of the tracking system. The innermost part of the new tracking system (called *Inner Tracker* or IT) is described elsewhere [3]; the *Outer Tracker* (OT) will be described in the following.

## 2. The CMS Outer Tracker for the High-Luminosity LHC

The CMS Outer Tracker will consist of four main large structures, the TB2S (Tracker Barrel with 2S modules), TBPS (Tracker Barrel with PS modules) and two TEDDs (Tracker Endcap Double Disks) on either side of the interaction region. Its proposed layout is shown in Fig. 1: The TB2S will consist of the outer three layers with modules parallel to the beam direction, the TBPS will be constituted by the inner three layers with progressively inclined modules. The TEDDs will complement the forward region with 5 double disks with modules perpendicular to the beam direction. The OT will consist of so-called $p_T$ *modules*, of which two basic types are foreseen, 2S and PS, which will be described in detail in Sec. 3. Both module types have two closely spaced silicon sensors, the 2S modules has two sensors with micro-strips, while the PS modules consist of one sensor with micro-strips and one sensor with macro-pixels. The most important feature of the modules which in turn is the defining feature of the OT is their ability to contribute information to the first CMS trigger stage (*Level-1* or *L1*). The basic paradigm is as follows: as a track passes through both sensors of a given module, the hit information from both sensors is correlated by one ASIC. As illustrated in Fig. 2 a track with high transverse moment ($p_T$) crosses both sensor planes with only a slight displacement in the precisely measuring coordinate. The readout ASIC defines a "window" which is used to decide whether a track satisfies the $p_T$ selection criterion. If it does, a local track segment, called *track stub*, is formed and subsequently pushed out to the Track Finder (cf. Sec. 4). Track stubs from all layers are assembled into tracks which will be used in the L1 decision. In order to enable a homogeneous $p_T$ selection threshold throughout the OT, different sensor spacings are used to change the lever arm for the track selection. For the 2S modules in the outer part spacings of 1.8 mm and 4.0 mm are used. For the PS modules three different spacings of 1.8 mm, 2.6 mm and 4.0 mm are employed.

The inclined geometry in the TBPS is employed for two main reasons. First, it allows to maintain hermetic coverage with fewer modules compared to a flat geometry. Secondly, the inclined geometry recovers stub inefficiencies due to tracks which cross only one of two sensors of a $p_T$ module or different halves of the same module for strongly inclined tracks (cf. [2, Fig. 2.6]).

The total active area of the CMS OT will be around 190 m$^2$

[*]Corresponding author
 *Email address:* erik.butz@kit.edu (Erik Butz)

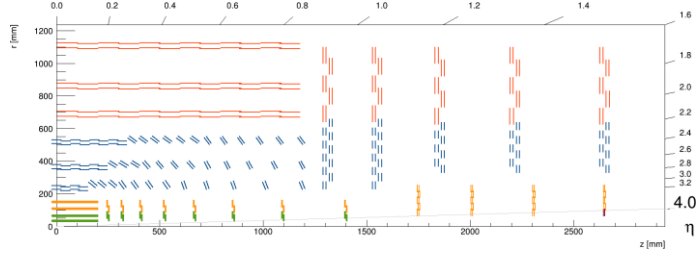

Figure 1: *rz* view of one quarter of the upgraded CMS inner tracking system for the HL-LHC. The Outer Tracker consists of the regions in blue and red (see main text for details).

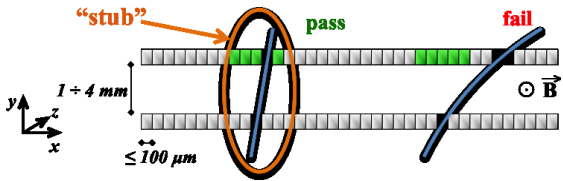

Figure 2: Track stub finding principle. A track passes both layers of a $p_T$ module. A high momentum track will be bent only slightly, passing through a selection window in the upper sensor prompting the readout ASIC to form and send out a *track stub*. A low momentum track falls outside the selection window and produces no stub.

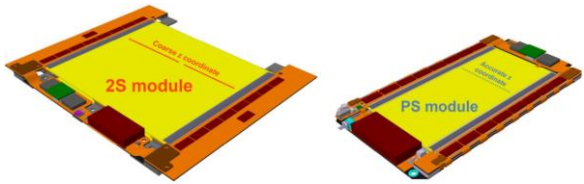

Figure 3: Schematic view of the two module types (left: 2S, right: PS) in the CMS Phase-2 Outer Tracker. The readout ASICs for the strip sensors are located on the readout hybrid on either side of the module. The ASICs for the pixelated sensor of the PS module (not shown, cf. main text for details) are bump-bonded to the sensor.

distributed over 13'296 individual modules.

## 3. Modules and Front-End Electronics

As stated in Sec. 2, the CMS OT will use two different basic module types, 2S and PS, which are shown in Fig. 3. Both module types have two different types of hybrids which house the readout and auxiliary electronics, respectively. The readout hybrid houses the readout ASICs and concentrator chip while the service hybrid[1] houses the opto-electronics and the DCDC converters for powering. There are two readout hybrids on either side of the module to house the two groups of readout ASICs (see below). For both module types the signals from the "upper"[2] and lower sensor need to be routed to one ASIC to perform the track stub finding. This is achieved by using a flexible hybrid which is folded over a spacer and which allows routing of the signal between the different parts of the module. A side view of the two module types with the ASICs and the flex-hybrid can be seen in Fig. 4.

*2S modules.* The two strip sensors in the 2S modules have a size of about $10 \times 10$ cm$^2$ and they are divided into two sets of 1016 strips with a cell size of 5 cm $\times$ 90 $\mu$m.

One ASIC is employed to read out both silicon sensors. The CMS Binary Chip (CBC) is implemented in 130 nm technology and reads out 254 strips, 127 from each of the two sensors of a 2S module. The CBC performs zero-suppression by applying a threshold cut on hits, performs the correlation of hits between the two sensors to form track stubs and pushes them out at bunch crossing rate. Two groups of 8 CBC chips each read out the two groups of micro-strips. They are located on the readout hybrid on either side of the module and are wire-bonded to the sensor. The individual CBC chips exchange data with their neighbors to enable stub finding across readout chip boundaries.

*PS modules.* The PS module consists of a strip sensor with a size of approximately $5 \times 10$ cm$^2$ with two sets of 960 strips with dimensions of 2.35 cm $\times$ 100 $\mu$m. The other sensor in the PS module is divided into macro-pixels of 1.5 mm length and pitch of 100 $\mu$m to match the pitch of the strip sensor.

For the PS modules there are two independent ASICs which readout the silicon sensors: the strip sensor is read out by the *Short Strip ASIC* (SSA) while the pixelated sensor is read out by the *Macro Pixel ASIC* (MPA). The track stub finding in the PS modules is done by the MPA. To this end, it receives the information about strip clusters from the SSA which sends them together with information about the bunch crossing in which a hit occurred. The MPA combines this information with the macro-pixel information to form track stubs. As for the 2S module there are two groups of 8 SSAs on each readout hybrid to read out the strip sensor. For the pixelated sensor, the readout ASICs is bump-bonded onto the sensor. Each MPA chip reads out 1'888 macro pixels, The macro-pixel sub-assembly (MaPSA) contains the actual silicon sensor and a total of 16 MPA readout chips bump-bonded to it.

*Concentrator ASIC.* For both module types the *Concentrator Integrated Circuit* (CIC), receives the data from the readout ASICs. Depending on the module type it performs data sparsification (2S) or deserialization (PS), formats the output data and sends them to the service hybrid (see below). The CIC needs to process both the track stub data (*TRIG data*) at 40 MHz and the detector payload (*DAQ data*) which gets pushed out after an L1 accept signal with a maximum rate of 750 kHz. The CIC receives data on six parallel lines at 320 Mb/s which corresponds to 48 bit/bx for each connected FE chip. The bandwidth is divided between TRIG and DAQ data with 5 bit bandwidth of TRIG and 1 bit of DAQ data sent every 3.125 ns.

---
[1]The 2S modules have one, the PS modules two service hybrids.
[2]*Upper* refers to the sensor visible in Fig. 3.

*Auxiliary Electronics.* The service hybrids house three main components needed for the operation of the modules in addition to the readout ASICs. In order to ship out the data read out by the readout ASICs and relayed by the CIC chips two components are used: The *low-power Gigabit Transceiver* (lpGBT) [4] together with the *Versatile Link+* (VL+) [5] receive the data transform them from an electrical to an optical signal and transmit them over optical fiber towards the backend electronics. The optical link is bi-directional and will also be used to transmit clock, trigger and fast-commands as well as configuration data to the module. The downstream link speed is either 5.12 or 10.24 Gb/s depending on the position in the detector and the expected particle occupancy while the upstream link speed is 2.56 Gb/s.

The Phase-2 OT will employ a multi-stage DCDC powering scheme based on DCDC converters. The input voltage provided by the power supply system on the experimental cavern balconies will be in the range 10–12 V. Inside the detector on the individual modules there are 2(2S modules) or 3(PS modules) DCDC converters to convert to the voltages used by the various module components. In a first step the voltage is transformed from 10–12 V down to 2.5 V which can be used directly for the biasing of the VCSELs. A second DCDC converter stage transforms the 2.5 V down to the required voltage for the ASICs (between 1.0 V and 1.25 V depending on ASIC and module type, cf. [2, p. 183]).

The modules of the OT are stand-alone units in the sense that they are directly connected to the backend electronics with no intermediary aggregator card of any kind.

## 4. Backend Electronics

The backend electronics consists of two parts, the DTC boards for the readout and control of the detector modules and the Track Finder for the further processing of the track stubs and subsequent forwarding to the other parts of the CMS L1 trigger system.

*Data, Trigger and Control board.* The main interlocutor for the modules of the OT is the *Data, Trigger and Control board* (DTC). It reads out both the TRIG and DAQ from the the front-end, relays clock, trigger and fast commands and performs the configuration of the detector modules. The DTC receives the data stream from the modules and separates the TRIG data stream from the DAQ data and forwards the former to the Track Finder (see below). The DTC board will be based on the *Advanced Telecommunications Computing Architecture* (ATCA) standard and each board will receive data from up to 72 detector modules.

*Track Finder.* The backend track finder system receives the stub data from the individual detector modules and performs track finding on them. The stub data gets pushed out from the detector at bunch crossing rate and for a pile-up of 200 it is estimated that there will be up to 15'000 track stubs per events. The track finding is performed in two steps, pattern recognition and track fitting. For both steps two different approaches are currently being considered. The Track Finder hardware will also be based on ACTA and will be using commercial FPGAs.

*Pattern Recognition.* In the pattern recognition track candidates are formed from the large number of track stubs. These track candidate are subsequently passed to the final track fit for best possible track parameter estimation in the L1 decision.

**Tracklet Approach** In the tracklet approach possible track candidates are identified starting from track segments in adjacent layers, called *tracklets*. The algorithm starts from seeding layers and checks if a tracklet can be formed from the two layers under consideration together with the primary interaction point as additional constraint. The algorithm then tries to find further hits with a simple road search algorithm by extrapolating the initial trajectory to other layers.

**Hough Transform** An alternative algorithm under consideration for the pattern finding is the *Hough transform*[6]. The general principle is to take all track stubs of the OT and apply the Hough transform on them which effectively projects them as straight lines into a 2-dimensional coordinate system. The lines of stubs which belong to a track candidates will intersect in one point (cf. Fig. 5). Track candidates which contain four or five track stubs will be considered for the final track fit.

*Track Fitting.* Irrespective of the initial pattern recognition method that will be employed, a final, refined track fit will be performed on the found track candidates in order to obtain the best possible estimate of the track parameters. The Kalman Filter approach [7] is commonly used for tracking applications in particle physics and a pipelined version of it is under consideration for the OT Track Finder. As an alternative, a simple linearized $\chi^2$ minimization is also being considered.

Hardware demonstrator systems for both approaches for pattern recognition and track fitting have been set up and show that tracks can be found with high efficiency and within the available time budget of about 3.2 $\mu$s.

## 5. Expected Performance

The upgrade CMS OT features a number of improvements compared to its predecessor and despite the vastly increased number of channels and increased power demands, the material budget of the upgraded tracker is expected to be significantly reduced. The main ingredients to achieve this are a reduced number of layers, optimized routing of the services, use of light weight material for support structures, low-mass $CO_2$ cooling and the use of DCDC converters. A comparison of the material budget for the phase-1 and phase-2 tracker is shown in Fig. 6. Also the performance for track reconstruction is better in basically all respects. In Fig. 7 the expected transverse momentum and impact parameter resolution is shown comparing the phase-1 CMS tracker and the upgraded system. It can be seen that the performance is both cases is significantly improved owing to the decreased cell size in the new system as well as the reduced scattering of the tracks due to the reduced material budget.

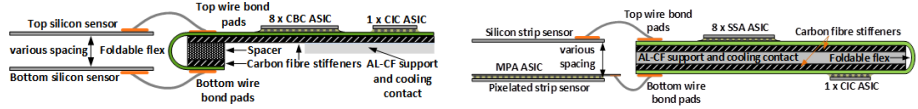

Figure 4: Cut-away side view of the readout hybrid of the 2S (top) and PS (bottom) modules with their respective ASICs. In both cases a flex hybrid is used to route the signals from the two sensors.

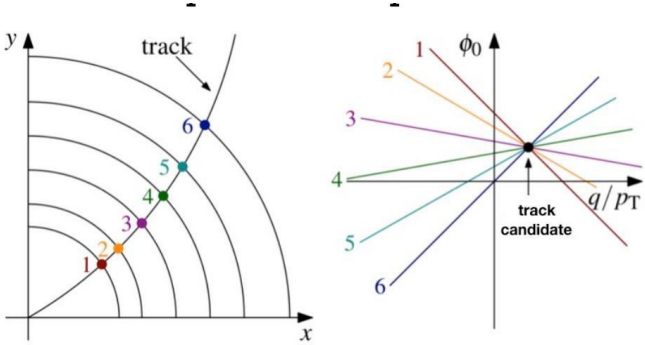

Figure 5: Illustration of the Hough transform principle which projects track stubs into the abstract $\phi_0$–$q/p_T$ space. Each track stub is represented by a straight line and track candidates are identified by intersecting lines from 4 or 5 track stubs.

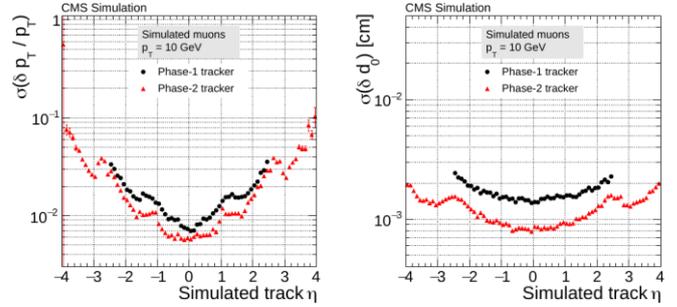

Figure 7: Comparison of track parameter resolution for the phase-1 CMS tracker (black) and the upgraded tracker (red). Both for the $p_T$ resolution (left) as well as the impact parameter resolution (right) the performance of the upgraded system is superior.

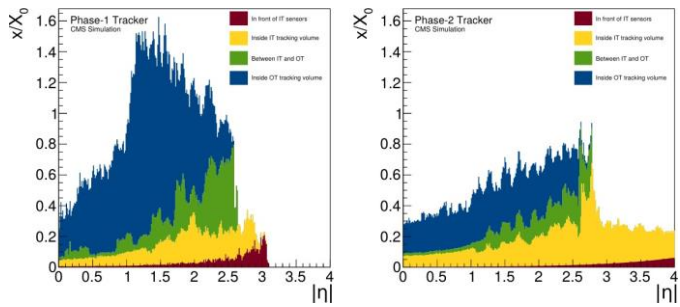

Figure 6: Material budget for the phase-1 CMS tracker (left) and the upgraded tracker (IT+OT) (right). Despite its increased number of readout channels, the upgraded OT material budget is greatly reduced compared to its predecessor.

## 6. Conclusions and Outlook

CMS is planning to completely replace its inner tracking system for the HL-LHC running. The central feature of the upgraded outer tracker is its ability to contribute tracks with $p_T > 2$ GeV as trigger primitives to the first level trigger decision. The trigger capability is provided by the $p_T$ modules used in the OT which foresee to enable correlation of hit information between the two closely spaced silicon sensors of the module. The modules are stand-alone units with all needed auxiliary electronics fully contained on the module. The powering of the system is achieved with multi-stage DCDC conversion. The backend track finding system is based on ACTA boards together with commercially available FPGAs. Two approaches for both pattern recognition and track fitting are under consideration and hardware demonstrators have been set up which show that the required performance can be reached.

The tracking performance of the upgraded system is expected to surpass the one of the currently installed and to be well-suited for the challenging environment posed by the HL-LHC environment.

## References

[1] G. Apollinari, I. Béjar Alonso, O. Brüning, P. Fessia, M. Lamont, L. Rossi, L. Tavian, High-Luminosity Large Hadron Collider (HL-LHC), CERN Yellow Rep. Monogr. 4 (2017) 1–516. doi:10.23731/CYRM-2017-004.

[2] CMS Collaboration, The Phase-2 Upgrade of the CMS Tracker, Tech. Rep. CERN-LHCC-2017-009. CMS-TDR-014, CERN, Geneva (Jun 2017). URL https://cds.cern.ch/record/2272264

[3] G. Squazzoni, The CMS Inner Tracker for the High Luminosity LHC, these Proceedings.

[4] CERN, LpGBT specification document. URL https://espace.cern.ch/GBT-Project/LpGBT/Specifications/LpGbtxSpecifications.pdf

[5] J. Troska, A. Brandon-Bravo, S. Detraz, A. Kraxner, L. Olanter, C. Scarcella, C. Sigaud, C. Soos, F. Vasey, The VTRx+, an optical link module for data transmission at HL-LHC, PoS TWEPP-17 (2017) 048. doi:10.22323/1.313.0048.

[6] P. V. C. Hough, Method and means for recognizing complex patterns, US Patent 3,069,654 (December 1962).

[7] R. Fruhwirth, Application of Kalman filtering to track and vertex fitting, Nucl. Instrum. Meth. A262 (1987) 444–450. doi:10.1016/0168-9002(87)90887-4.